\documentclass[conference]{IEEEtran}
\IEEEoverridecommandlockouts
\usepackage{cite}
\usepackage{amsmath,amssymb,amsfonts}
\usepackage{algorithmic}
\usepackage{graphicx}
\usepackage{textcomp}
\usepackage{xcolor}
\def\BibTeX{{\rm B\kern-.05em{\sc i\kern-.025em b}\kern-.08em
    T\kern-.1667em\lower.7ex\hbox{E}\kern-.125emX}}
\usepackage{tikz} 

\newcommand{\mycopyrightnotice}{%
  \begin{tikzpicture}[remember picture,overlay]
    \node[anchor=south, yshift=0.4in] at (current page.south) {%
      \parbox{\textwidth}{\fontsize{7.5pt}{9pt}\selectfont
      \copyright~2026 IEEE. Personal use of this material is permitted. Permission from IEEE must be obtained for all other uses, in any current or future media, including reprinting/republishing this material for advertising or promotional purposes, creating new collective works, for resale or redistribution to servers or lists, or reuse of any copyrighted component of this work in other works.\\
      Accepted for publication in the Proc. 48th Annu. Int. Conf. IEEE EMBS (EMBC 2026), Toronto, Canada, July 20-24, 2026.}
    };
  \end{tikzpicture}%
}

\begin{document}

\title{End-to-End Machine Learning for Depressive State Classification via EEG and fNIRS}

\author{\IEEEauthorblockN{Riki Sakurai$^\circledast$}
\IEEEauthorblockA{\textit{Graduate School of Engineering and Science} \\
\textit{Shibaura Institute of Technology,} Tokyo, Japan \\
\textit{FINDEX Inc.,} Tokyo, Japan\\
\textit{RIKEN AIP,} Tokyo, Japan\\
ORCID:0009-0006-3576-845X}
\and
\IEEEauthorblockN{Simon Kojima\quad\quad\quad\quad\quad\quad}
\IEEEauthorblockA{\textit{Inria Centre at the University of Bordeaux\quad\quad\quad\quad\quad\quad} \\
Talence, France\quad\quad\quad\quad\quad\quad \\
ORCID: 0000-0002-6509-7363\quad\quad\quad\quad\quad\quad}
\and
\IEEEauthorblockN{Mihoko Otake-Matsuura}
\IEEEauthorblockA{\textit{RIKEN AIP,} Tokyo, Japan \\
ORCID: 0000-0003-3644-276X}
\and
\IEEEauthorblockN{Shin'ichiro Kanoh$^\divideontimes$}
\IEEEauthorblockA{\textit{College of Engineering} \\
\textit{Shibaura Institute of Technology} 
\\Tokyo, Japan \\
\textit{RIKEN AIP,} Tokyo, Japan\\
ORCID: 0000-0002-6554-5648}
\and
\IEEEauthorblockN{Tomasz M. Rutkowski$^{\divideontimes,\circledast}$\thanks{$^\divideontimes$Coresponding author.}\thanks{$^\circledast$At the time of the data collection, the author was affiliated with RIKEN AIP, Tokyo, Japan.}}
\IEEEauthorblockA{
\textit{The University of Tokyo,} Tokyo, Japan \\
\textit{Nicolaus Copernicus University}, Torun, Poland \\
\textit{RIKEN AIP,} Tokyo, Japan\\
\textit{Araya Inc.}, Tokyo, Japan \\
ORCID: 0000-0002-4259-4121}
}

\maketitle

\mycopyrightnotice 

\begin{abstract}
The escalating demand for mental healthcare, driven by rising societal stress, highlights the limitations of traditional psychiatric diagnostics. Conventional methods—relying primarily on clinical interviews and patient self-reports—are inherently vulnerable to subjective bias and the varying empirical judgment of practitioners. To address the need for quantitative evaluation, biological signal-based detection, including electroencephalography (EEG) and functional near-infrared spectroscopy (fNIRS), has emerged as a promising objective alternative. Such technology is particularly vital for identifying latent depressive states that may be unrecognized by the subjects themselves. Furthermore, in aging populations, the high comorbidity between depression and dementia necessitates early differentiation to prevent mutual symptom exacerbation and maintain Quality of Life (QoL). 
This pilot study of eleven healthy students establishes a framework for biological signal-based depression detection, serving as a foundational step toward automated, objective diagnostic tools for clinical use.
\end{abstract}


\section{Introduction}

In recent years, the global demand for mental healthcare has escalated significantly, driven by rising societal stressors. Traditional diagnostic frameworks for psychiatric disorders and psychological states rely heavily on clinical interviews and patient self-reports~\cite{beck1996beck}. However, these conventional methods lack quantitative precision, as they are inherently influenced by the patient’s subjective perception and the clinician's empirical judgment.
To address these limitations, the detection of psychological states through biological signals—specifically electroencephalography (EEG) and functional near-infrared spectroscopy (fNIRS)—has emerged as a robust, objective alternative~\cite{tomekFN2023,tomekFAN2024,tomekEMBC2024_2}. EEG offers high temporal resolution, capturing rapid neural oscillations, while fNIRS provide superior spatial insights into cortical hemodynamic responses. The integration of these two modalities allows for a synergistic and comprehensive evaluation of mental states that is less susceptible to conscious bias than traditional assessments~\cite{elnaggar2025}.
Early intervention in depression is critical to preventing disease progression. Biological signal-based screening has the potential to identify latent depressive markers of which the subjects may remain unaware. In particular, deficits in emotional working memory—the cognitive ability to maintain and process affective information—are recognized as significant neurophysiological indicators of depressive tendencies.
In this preliminary study, we propose a framework for the early detection of depressive states by analyzing neural responses during an emotional short-term memory task. We conducted a simultaneous EEG-fNIRS experiment where participants performed tasks requiring the retention of emotional stimuli. These multimodal signals were labeled using the Beck Depression Inventory (BDI) scores. 
Finally, as a preliminary proof-of-concept, we utilize SincShallowNet~\cite{ravanelli2018speaker,schirrmeister2017deep,braindecode} to evaluate the feasibility of estimating depression levels from the neural dynamics of a still predominantly healthy cohort. 

\section{Methods}

Eleven healthy university students (5 female, 6 male) participated in this study, with a mean age of $22.27 \pm 1.21$ years. The experiment was conducted at the RIKEN Center for Advanced Intelligence Project (AIP) in Tokyo, Japan. All procedures were approved by the RIKEN Ethical Committee for Experiments with Human Subjects (Permission No. Wako3 30-28(4)) and were performed in accordance with the Declaration of Helsinki. Electroencephalography (EEG) and functional near-infrared spectroscopy (fNIRS) data were recorded simultaneously for all participants.
\begin{figure}[t]
\centerline{\includegraphics[height=5cm]{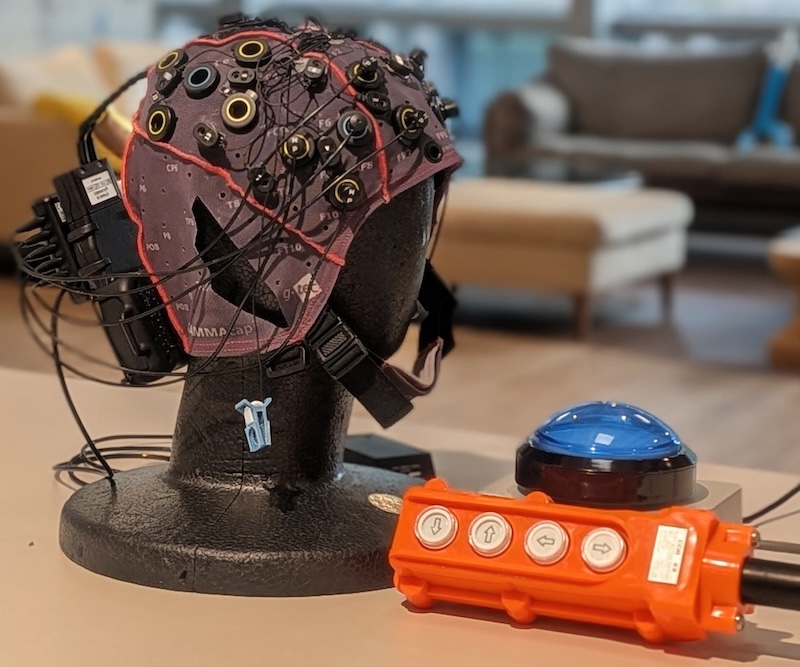}}
\caption{Experimental hardware and response interface. The g.NAUTILUS~fNIRS system (g.tec medical engineering GmbH, Austria) shown alongside the behavioral response buttons. The orange panel is utilized for target stimulus selection, while the large blue button facilitates self-paced progression to the next trial within the affective stimulus paradigm.}
\label{fig:equipment}
\end{figure}

Prior to the measurements, participants completed the Beck Depression Inventory-II (BDI-II)~\cite{beck1996beck}. While the original inventory consists of $21$ items, a $20-$item version was utilized; the item concerning sexual interest was omitted to minimize participant discomfort and address potential privacy concerns.
\begin{table}[b]
    \centering
    \caption{Ranges of Valence and Arousal Scores for Each Difficulty}
    \label{tab:difficulty_level}
    \begin{tabular}{|c||c|c|}
        \hline
        Difficulty  & Valence Range                 & Arousal Range \\
        \hline \hline
        Easy        & $[-1.0,-0.5]$ or  $[0.5,1.0]$ & $[-1.0,-0.5]$ or  $[0.5,1.0]$\\
        Middle & 
         $[-0.5,-0.3]$ or  $[0.3,0.5]$ & $[-0.5,-0.3]$ or  $[0.3,0.5]$ \\
        Hard & 
        $[-0.3,0.3]$ & $[-0.3,0.3]$ \\
        \hline
    \end{tabular}
\end{table}

\subsection{Experimental Protocol}\label{sec:methodEXP}

To elicit neural responses associated with potential depression biomarkers, we designed a novel working memory paradigm. This task requires subjects to memorize emotional expressions presented as short facial videos or spoken auditory utterances, sourced from the ``Mind Reading'' database~\cite{mindREADING}. 
The experiment consisted of six sessions in total: three difficulty levels for each of the two stimulus modalities (video and audio). Each session comprised ten trials. To ensure high-quality EEG data, subjects were instructed to provide their responses only after the full presentation of the affective stimuli, thereby minimizing muscle activity interference with the recording.

\subsubsection{Stimulus Selection and Difficulty Levels}

Stimuli were selected based on the NRC VAD Lexicon~\cite{mohammad2025}, which provides valence, arousal, and dominance scores for over $50,000$ English words on a scale from $-1$ to $1$. We established three difficulty levels by adjusting the required valence and arousal ranges for stimulus selection (see Table~\ref{tab:difficulty_level}). We hypothesized that larger absolute values in these ranges (high intensity) would facilitate easier emotion discrimination, while values closer to zero (subtle intensity) would increase task difficulty.

The selection process followed a specific hierarchy:
\begin{itemize}
    \item {\bf Target Selection:} An emotional word was randomly selected from a pool meeting the specific valence/arousal range for that difficulty level. 
    \item {\bf Distractor Selection:} A ``non-target'' word was chosen from a pool with a valence sign opposite to that of the target word.
    \item {\bf Media Allocation:} Three media files were extracted from the database: two representing the target emotion (one for the initial ``target'' phase and one for the ``comparison'' phase) and one representing the non-target emotion.
\end{itemize}
\begin{figure}[t]
    \centerline{\includegraphics[height=5cm]{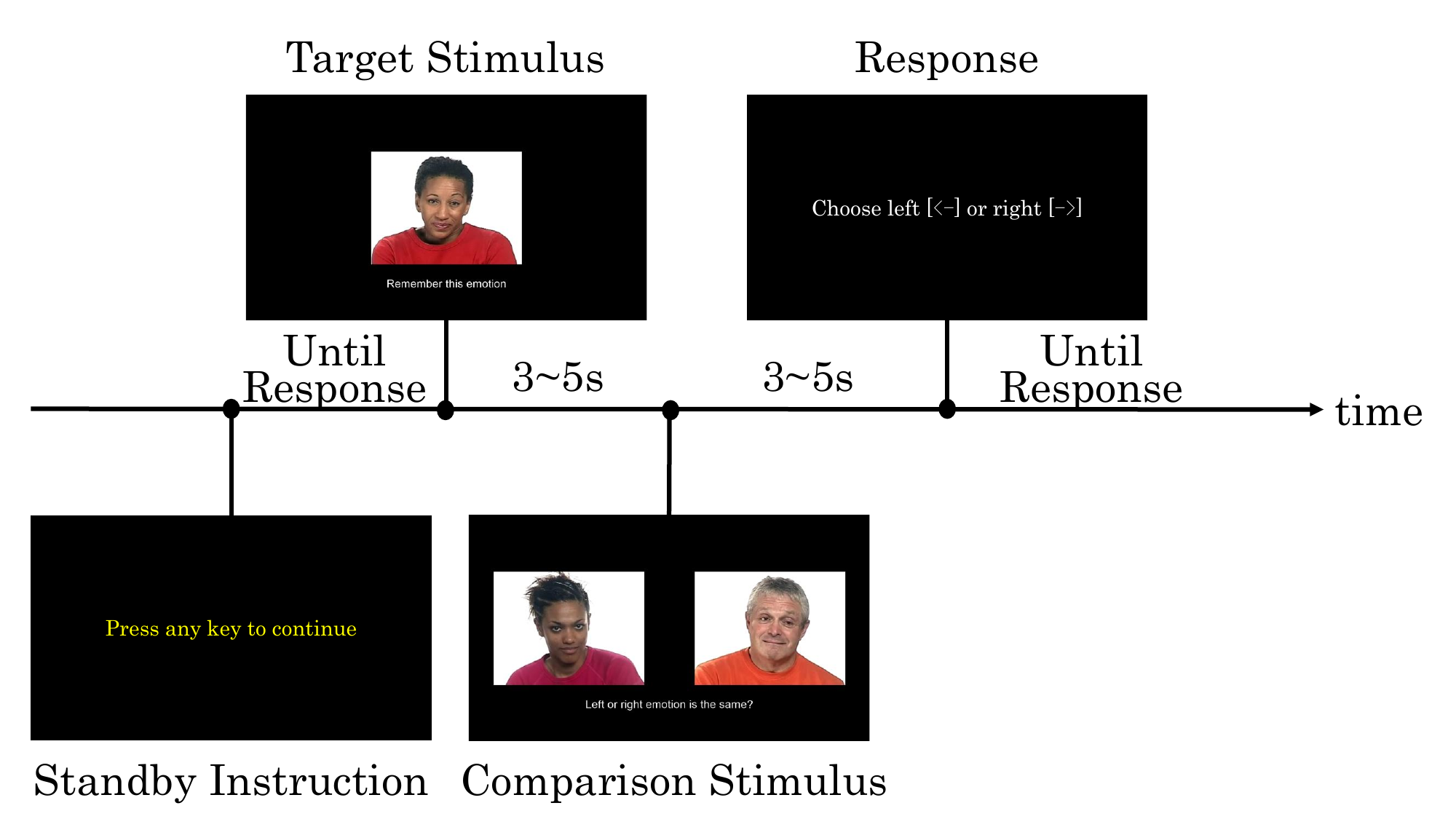}}
    \caption{Experimental procedure for the emotional video modality. In the auditory modelity (empitional speech atterances, instead of video clips, the spoken stimuli were delivered via headphones). }
    \label{fig:experiment}
\end{figure}
    
\subsubsection{Experimental Procedure}
    
Each trial followed a structured sequence:
\begin{itemize}
    \item {\bf Target Phase:} Subjects were presented with a ``target stimulus''—either a five-second video clip of a professional actor or an audio recording of affective speech.
    \item {\bf Maintenance Phase (Working Memory):} Subjects held the emotional state in memory.
    \item {\bf Test Phase:} Two new affective stimuli were presented simultaneously. One stimulus matched the emotion of the target, while the other was a non-matching (distractor) expression.
        \begin{itemize}
            \item {\bf Video Condition:} Two videos were displayed side-by-side with synchronized start times.
            \item {\bf Audio Condition:} Stimuli were presented dichotically. To prevent immediate recognition via speech onset, the onset of one channel (randomly assigned left or right) was delayed by one second.
        \end{itemize}
    \item {\bf Subject Response:} Participants utilized a specialized interface to perform a forced-choice task, indicating their selection by pressing the left or right arrow keys located within the orange response box (see Figure~\ref{fig:equipment}).
    \item {\bf Self-Paced Protocol:} To minimize fatigue and ensure attentional readiness, a self-pacing mechanism was employed; subjects initiated subsequent trials at their discretion by engaging a prominent blue trigger button (Figure~\ref{fig:equipment}).
\end{itemize}    
The experimental flow for a single trial within the video condition is illustrated in Figure~\ref{fig:experiment}. To ensure precise temporal alignment with the EEG recordings, all experimental triggers and event markers were captured using the Lab Streaming Layer (LSL) protocol~\cite{lsl}.

\subsection{EEG and fNRIS Recording and Preprocessing}\label{sec:methodPREPROC}

\begin{figure}[t]
\centerline{\includegraphics[width=0.9\columnwidth]{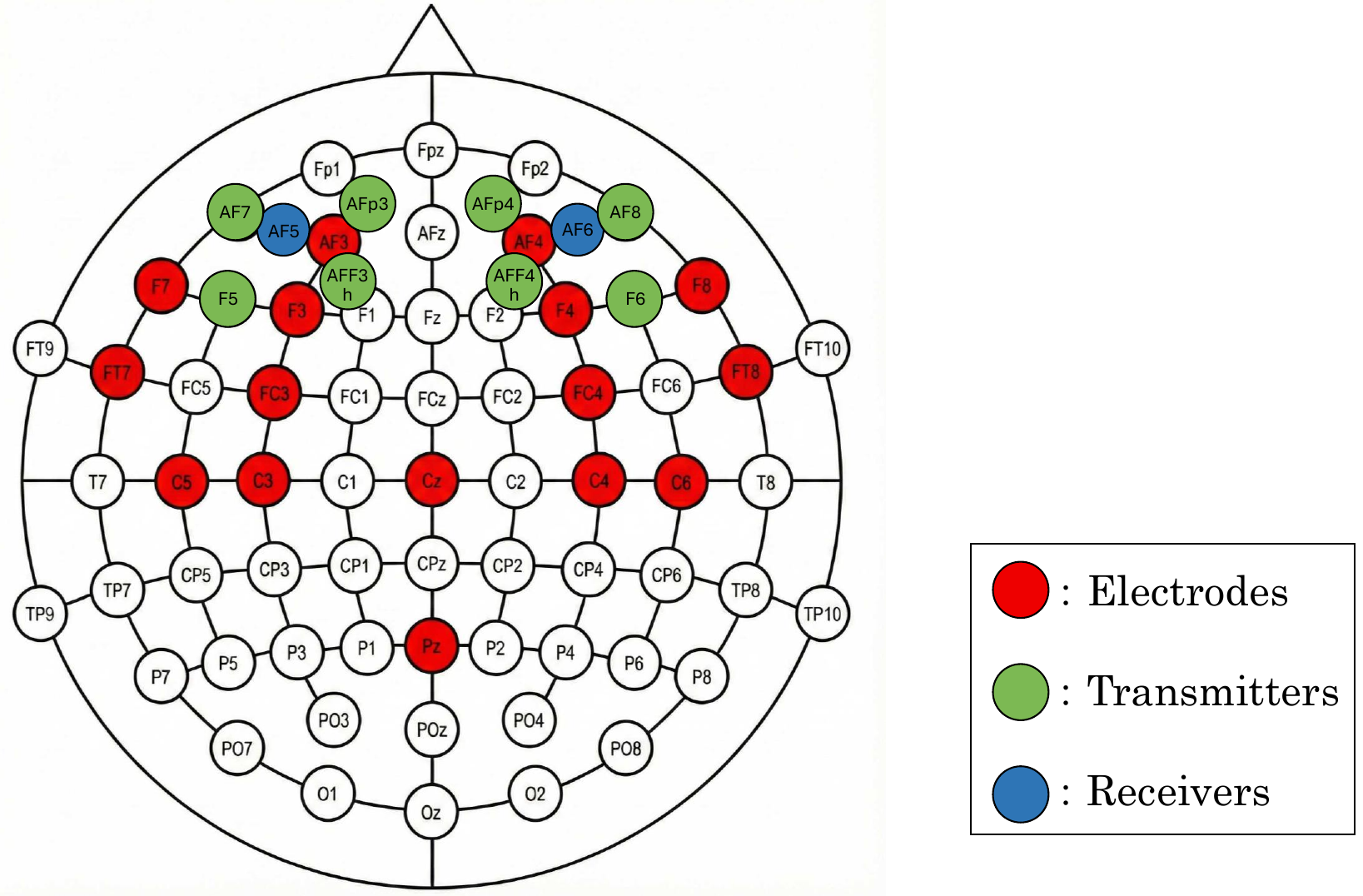}}
\caption{Hybrid EEG-fNIRS sensor topography. Schematic representation of the optode and electrode montage for the g.NAUTILUS~fNIRS wireless system (g.tec medical engineering GmbH, Austria). EEG electrode sites are denoted by red markers, while fNIRS light sources (transmitters) and detectors (receivers) are represented by green and blue markers, respectively, illustrating the integrated spatial distribution across the scalp.}
\label{fig:cap}
\end{figure}

Simultaneous EEG and fNIRS recordings were conducted using a multimodal g.NAUTILUS~fNIRS system (g.tec medical engineering GmbH, Austria). The spatial distribution of the hybrid sensor montage is illustrated in Figure~\ref{fig:cap}.

\subsubsection{EEG Configuration} Sixteen active EEG electrodes were positioned at AF3, AF4, F7, F8, F3, F4, FC3, FC4, C5, FT7, Cz, FT8, C6, C3, C4, and Pz, following the International $10-10$~system. The reference was placed on the right earlobe, with the ground at AFz. During acquisition, data were hardware-filtered with a $1$~Hz high-pass filter and a $50$~Hz notch filter to mitigate power-line interference.

\subsubsection{fNIRS Configuration}
Cortical hemodynamic activity within the frontal lobe was monitored via an eight-channel optical subsystem integrated into the g.NAUTILUS~fNIRS platform. The optode arrangement consisted of eight sources (transmitters at $AFp4$, $AF8$, $F6$, $AFF4h$, $AF7$, $AFp3$, $AFF3h$, $F5$, positions) and two detectors (receivers at $AF6$, $AF5$, positions), as illustrated in Figure~\ref{fig:cap}. To ensure sufficient cortical penetration while maintaining signal integrity, a fixed source-detector separation of $32$~mm was utilized for all channels.

\subsubsection{Preprocessing Pipeline} Multimodal signals were synchronously digitized at a sampling frequency of $250$~Hz. For each experimental trial, a three-second epoch was extracted, precisely time-locked to the stimulus onset to unify the temporal alignment of stimuli with varying durations ($3\sim 5'$~s). 
The data were processed as follows:
\begin{itemize}
    \item {\bf EEG Preprocessing:} To isolate neurophysiological oscillations and attenuate low-frequency drifts and high-frequency interference, a fourth-order Butterworth bandpass filter ($3$--$45$~Hz) was applied.
    \item {\bf fNIRS Preprocessing:} Raw optical intensity fluctuations were acquired at $760$~nm and $850$~nm. These signals were converted to optical density changes and subjected to a $0.5$~Hz low-pass filter to eliminate high-frequency noise and pulsatile cardiac artifacts. The resulting cleaned signals served as the primary features for the downstream classification architecture.
\end{itemize} 

\subsection{Subject Group Definition and Labeling}\label{sec:methodLABELING}

To simulate a binary screening scenario, participants were categorized into two groups based on their depressive tendencies as measured by the Beck Depression Inventory (BDI-II)~\cite{beck1996beck}. A cutoff score of $7$ was adopted to achieve a balanced data distribution for the classification task. Accordingly, subjects scoring $\leq 6$ were assigned to the Low-Score (LS) group, while those scoring $\geq 7$ were assigned to the High-Score (HS) group.

We acknowledge that the HS group in this study largely falls within the "minimal depression" range according to standard clinical thresholds~\cite{beck1996beck}. However, this specific criterion was selected to investigate subtle neurophysiological variations in psychological tendencies within a predominantly healthy cohort, reflecting a ``sub-clinical'' screening objective. For the deep learning classification task, the LS and HS groups were assigned binary labels of $0$ and $1$, respectively.

\subsection{End-to-end Machine Learning}\label{sec:methodML}

To classify depressive states from the multimodal dataset—comprising $16$~EEG channels and $8$~fNIRS optical density signals—we employed SincShallowNet~\cite{ravanelli2018speaker}, implemented via the Braindecode framework~\cite{schirrmeister2017deep,braindecode}. This architecture is a hybrid variant of the ShallowConvNet~\cite{schirrmeister2017deep}, specifically optimized for raw signal decoding by integrating parameterized Sinc-filters originally proposed for acoustic waveform processing~\cite{ravanelli2018speaker}.

\noindent The model’s pipeline consists of three primary functional stages:
\begin{enumerate}
    \item {\bf Parameterized Temporal Filtering:} Unlike standard Convolutional Neural Networks (CNNs) that learn arbitrary kernel weights, the initial layer utilizes Sinc-functions to perform band-pass filtering. Each filter is defined solely by learnable low ($f_1$) and high ($f_2$) cutoff frequencies. This constraint ensures the extraction of physically meaningful frequency features (e.g., $\alpha$/$\theta$ oscillations in EEG or hemodynamic trends in fNIRS) while significantly reducing parameter complexity and the risk of overfitting.
    \item {\bf Multimodal Spatial Integration:} Following temporal feature extraction, a spatial depthwise convolution is applied across the combined $24-$channel input ($16$~EEG $+$ $8$~fNIRS). This layer performs a weighted integration of the electrical and hemodynamic features, effectively learning the cross-modal spatial correlations necessary for identifying latent depressive indicators across different cortical regions.
    \item {\bf Feature Transformation and Classification:} To capture signal power distributions, the architecture applies a square activation function followed by mean pooling and a logarithmic transformation. This sequence approximates the log-variance of the signals, which serves as a robust feature for identifying state-dependent changes in neural activity. The resulting feature vector is mapped to the final classification output through a fully connected layer.
\end{enumerate}
We implemented the SincShallowNet architecture using the Braindecode framework~\cite{schirrmeister2017deep,braindecode}. This model integrates the hierarchical spatial-temporal features of ShallowConvNet [1] with the parameterized sinc-filters originally proposed for raw waveform processing~\cite{ravanelli2018speaker}.

\section{Results}

\begin{figure}[t]
\centerline{\includegraphics[width=0.9\columnwidth]{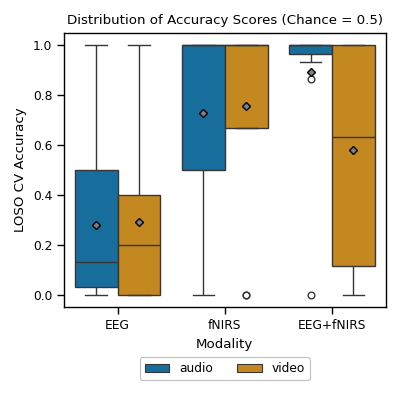}}
\vspace{-0.1cm}
\caption{Leave-one-subject-out cross-validation results. Comparison of classification accuracy for depressive versus normal mental states across three signal modalities (EEG, fNIRS, and combined) and two affective stimuli (audio vs. video). Horizontal lines represent medians, while dots indicate means.}
\label{fig:results}
\end{figure}

We evaluated the performance of SincShallowNet~\cite{ravanelli2018speaker} on multimodal EEG-fNIRS data using the Braindecode framework~\cite{schirrmeister2017deep}. Figure~\ref{fig:results} summarizes these results, highlighting superior predictive power in the auditory affective speech condition. In this modality, the integration of EEG and fNIRS for depressive trait prediction achieved peak performance, with central tendencies (mean and median) consistently surpassing the $0.80$ accuracy threshold.

\section{Conclusions}

This study demonstrates the efficacy of an end-to-end deep learning approach for the objective identification of depressive traits using multimodal biological signals. By leveraging the SincShallowNet architecture within the Braindecode framework, we successfully integrated EEG and fNIRS data to overcome the inherent subjectivity of traditional psychiatric assessments.
Our preliminary results indicate that:
\begin{itemize}
    \item {\bf Multimodal Integration:} The combination of EEG and fNIRS provides a robust physiological profile, achieving peak predictive performance particularly during auditory affective speech tasks.
    \item {\bf Performance Benchmarks:} The model achieved high classification reliability, with mean and median accuracies consistently exceeding the $0.80$ threshold, significantly outperforming the chance level of $0.50.$
    \item {\bf Clinical Implications:} These findings validate the potential of automated, biological-signal-based tools as early screening mechanisms for latent depression. Such objective metrics are crucial for differentiating depression from comorbid conditions like dementia in aging populations and for identifying sub-clinical states in younger cohorts.
\end{itemize}
While these preliminary results are limited by a small, healthy student cohort, they establish a foundation for future work, which will focus on validating the framework within larger clinical populations to refine its diagnostic accuracy and longitudinal stability.


\begin{thebibliography}{10}

\bibitem{beck1996beck}
A.~T. Beck, R.~A. Steer, and G.~Brown, ``Beck depression inventory--{II},'' {\em Psychological assessment}, 1996.

\bibitem{tomekFN2023}
T.~M. Rutkowski, M.~S. Abe, T.~Komendzinski, H.~Sugimoto, S.~Narebski, and M.~Otake-Matsuura, ``Machine learning approach for early onset dementia neurobiomarker using {EEG} network topology features,'' {\em Frontiers in Human Neuroscience}, vol.~17, p.~1155194, 2023.

\bibitem{tomekFAN2024}
T.~M. Rutkowski, T.~Komendzinski, and M.~Otake-Matsuura, ``Mild cognitive impairment prediction and cognitive score regression in the elderly using {EEG} topological data analysis and machine learning with awareness assessed in affective reminiscent paradigm,'' {\em Frontiers in Aging Neuroscience}, vol.~15, p.~1294139, 2024.

\bibitem{tomekEMBC2024_2}
S.~Kojima, R.~Shiba, Y.~Morimoto, K.~Furukawa, S.~Kanoh, T.~Komendzi{\'n}ski, M.~Otake-Matsuura, and T.~M. Rutkowski, ``Spatial auditory soundscapes for developing digital neurobiomarkers or cognitive interventions in early-onset dementia based on {EEG} and {fNIRS} machine-learning analysis,'' in {\em 46th Annual International Conference of the IEEE Engineering in Medicine \& Biology Society (EMBC)}, (Orlando, USA), pp.~1, poster ID 8331, IEEE Press, 2024.

\bibitem{elnaggar2025}
{K. Elnaggar, M. M. El-Gayar and M. Elmogy}, ``{Depression Detection and Diagnosis Based on Electroencephalogram ({EEG}) Analysis: A Systematic Review},'' {\em Diagnostics (Basel)}, vol.~15, no.~2, p.~210, 2025.

\bibitem{ravanelli2018speaker}
M.~Ravanelli and Y.~Bengio, ``Speaker recognition from raw waveform with {SincNet},'' in {\em 2018 IEEE Spoken Language Technology Workshop (SLT)}, pp.~1021--1028, IEEE, 2018.

\bibitem{schirrmeister2017deep}
R.~T. Schirrmeister, J.~T. Springenberg, L.~D.~J. Fiederer, M.~Glasstetter, K.~Eggensperger, M.~Tangermann, F.~Hutter, W.~Burgard, and T.~Ball, ``Deep learning with convolutional neural networks for {EEG} decoding and visualization,'' {\em Human Brain Mapping}, vol.~38, no.~11, pp.~5391--5420, 2017.

\bibitem{braindecode}
B.~Aristimunha, P.~Guetschel, M.~Wimpff, L.~Gemein, C.~Rommel, H.~Banville, M.~Sliwowski, D.~Wilson, S.~Brandt, T.~Gnassounou, J.~Paillard, B.~{Junqueira Lopes}, S.~Sedlar, T.~Moreau, S.~Chevallier, A.~Gramfort, and R.~T. Schirrmeister, ``Braindecode: toolbox for decoding raw electrophysiological brain data with deep learning models.''

\bibitem{mindREADING}
S.~Baron-Cohen, {\em {Mind Reading: The Interactive Guide to Emotions}}.
\newblock London, UK: Jessica Kingsley Publishers, 2004.

\bibitem{mohammad2025}
{S. M. Mohammad}, ``{NRC VAD Lexicon v2:} norms for valence, arousal, and dominance for over 55k {English} terms,'' {\em arXiv preprint arXiv:2503.23547}, 2025.

\bibitem{lsl}
T.~Stenner, C.~Boulay, {\em et~al.}, ``sccn/liblsl: v1. 16.2,'' {\em Zenodo}, 2022.

\end{thebibliography}

\addtolength{\textheight}{-12cm}   

\end{document}